\documentclass[twocolumn,preprintnumbers,amsmath,amssymb,floatfix,prl]{revtex4-1}
\usepackage{graphicx}
\usepackage{dcolumn}
\usepackage{bm}
\usepackage{ulem}
\usepackage[usenames,dvipsnames]{color}% use as \textcolor{red}{******} 

\begin{document}

\title{Typicality-Based Variational Cluster Approach to Thermodynamic Properties \\of the Hubbard Model}

\author{Hisao Nishida$^1$}
\author{Ryo Fujiuchi$^{1}$}
\author{Koudai Sugimoto$^2$}
\author{Yukinori Ohta$^1$}

\affiliation{$^1$Department of Physics, Chiba University, Chiba 263-8522, Japan}
\affiliation{$^2$Department of Physics, Keio University, Yokohama 223-8522, Japan}

\date{\today}

\begin{abstract}
We develop the variational-cluster-approximation method based on the 
thermal-pure-quantum-state approach and apply the method to the calculations 
of the thermodynamic properties of the Hubbard model, thereby obtaining the temperature 
dependence of the single-particle excitation spectra, entropy and specific heat, and 
order parameter of the antiferromagnetic phase efficiently.  We thus demonstrate 
that the method developed here has a potential power for elucidating finite-temperature 
properties of strongly correlated electron systems. 
\end{abstract}

%\pacs{}

\maketitle

%%%%%%%%%%%%%%%
%\section{Introduction}
%%%%%%%%%%%%%%%

Calculation of finite-temperature properties of strongly correlated quantum systems, 
for which perturbation theory breaks down, is a hard and challenging task because 
many excited states must be taken into account for evaluation of their physical quantities.  
In the grand canonical theory, we need to know all the eigenvalues $E_{\nu}$ and 
corresponding eigenstates $| \nu \rangle$ of the relevant Hamiltonian to evaluate 
the ensemble average of an operator $\hat{O}$ at temperature $T$, which is given as  
\begin{equation}
 \langle \hat{O} \rangle^{\mathrm{ens}}_{\beta, \mu}
	= \frac{1}{Z (\beta, \mu)} \sum_{\nu} e^{-\beta \left( E_\nu - \mu N_{\nu} \right)} \langle \nu | \hat{O} | \nu \rangle,
\label{eq:grand_canonical_ens}
\end{equation}
where $\beta = 1/T$ is the reciprocal temperature, $\mu$ is the chemical potential, 
$Z (\beta, \mu) = \sum_{\nu} e^{-\beta \left( E_\nu - \mu N_\nu \right)}$ is the grand 
partition function, and $N_\nu$ is the number of particles of the state $| \nu \rangle$.  
A straightforward way to solve this problem is to fully or partially diagonalize the 
Hamiltonian.  Although for quantum lattice models one can in principle evaluate this 
average if the system size is small enough, one immediately encounters a severe 
restriction because the dimension of the Hilbert space grows rapidly as the system 
size increases. 

In this respect, quantum cluster approximations based on the self-energy functional 
theory \cite{Maier2005RMP, Senechal0806arXiv} are found to be useful for treating 
the systems in the thermodynamic limit, the techniques of which include 
the dynamical cluster approximation, 
the single-site or cellular dynamical-mean-field theory (DMFT), 
the cluster perturbation theory (CPT), and 
the variational cluster approach (VCA).  
The DMFT combined with the quantum Monte Calro (QMC) method 
for a single-impurity solver \cite{Maier2005RMP, Georges1996RMP, Gull2011RMP} 
is often used to obtain the electronic states at finite temperatures.  However, it is 
well-known that the QMC method involves the notorious negative-sign problem 
\cite{Loh1990PRB, Troyer2005PRL}, especially in itinerant fermionic systems.  

Recently, the thermal-pure-quantum (TPQ) state approach based on the typicality 
has been found to be useful for obtaining physical quantities at finite temperatures 
\cite{Sugiura2012PRL, Sugiura2013PRL, Hyuga2014PRB}.  
Based on the typicality, i.e., the fact that almost all the pure quantum states 
on an energy shell are equilibrium states 
\cite{Goldstein2006PRL, Popescu2006NP, Sugita2007NPCS, Reimann2007PRL}, 
one can construct the TPQ state by lowering temperature, starting from a 
random vector corresponding to a thermal equilibrium state at infinite temperature 
\cite{Sugiura2012PRL, Sugiura2013PRL, Hyuga2014PRB}.  Some pioneering work 
\cite{Imada1986JPSJ, Hams2000PRE, Jaklic1994PRB, Jaklic2000AP, Aichhorn2003PRB} 
also uses the random vector to obtain the thermal average.  
Thus, the ensemble average at a finite temperature may be calculated by a single 
TPQ state obtained from the imaginary-time evolution of the random vector, instead 
of calculating all the relevant eigenstates and summing over the average values 
multiplied by the Boltzmann weight, as in Eq.~(\ref{eq:grand_canonical_ens}).  
Thereby, the computational cost can be reduced drastically in this approach although 
one needs to take a random-sampling average over the TPQ states.  
Another advantage of this approach is that there are no negative-sign problems 
occurring in the QMC method.  Moreover, because one needs not diagonalize 
the huge Hamiltonian matrix, one can treat the larger-size systems.  
We should also emphasize that one can obtain not only the static thermal quantities 
but also the dynamical quantities such as the excitation spectrum using the 
typicality-based analyses \cite{Bartsch2009PRL, Elsayed2013PRL, Steinigeweg2014PRL, 
Monnai2014JPSJ, Jin2015PRB, Steinigeweg2016PRB, Yamaji2018arXiv, Richter2019PRB}. 

In this paper, we calculate the finite-temperature properties of the Hubbard model, 
a representative of strongly correlated electron models, by making use of the 
TPQ-state approach.  
First, we use this approach in the CPT \cite{Senechal2000PRL, Senechal2002PRB} 
and calculate the single-particle excitation spectra in the thermodynamic limit.  
We then use this approach in the VCA 
\cite{Potthoff2003EPJB, Potthoff2003EPJB2, Potthoff2003PRL, Dahnken2004PRB} 
and calculate the grand potential as a function of the symmetry-breaking Weiss field 
for antiferromagnetism.  We moreover calculate the static thermal quantities, such 
as the entropy and specific heat of the system.  
In a previous study \cite{Seki2018PRB}, the same model as ours was investigated 
by the VCA, where the block Lanczos method was used as a quantum cluster solver.  
There, the calculations of the single-particle Green's functions for all eigenstates 
with nonvanishing Boltzmann weight are required, and therefore the computational 
cost rapidly rises up at high temperatures.  
On the other hand, the method presented here requires the Green's functions 
for a single TPQ state, thereby reducing the computational cost drastically. 
We thus demonstrate that the TPQ-state approach provides a powerful technique 
for elucidating the thermodynamic properties of the Hubbard model even at high 
temperatures.  

In what follows, we first introduce the TPQ-state formalism briefly, and then present 
the results obtained by the CPT and VCA combined with the TPQ-state approach.  
We consider the single-band Hubbard model defined by the Hamiltonian 
\begin{equation}
 \hat{\mathcal{H}} = -t \sum_{ \langle i, j \rangle} \hat{c}^\dagger_{i, \sigma} \hat{c}_{j, \sigma}
	+ U \sum_{i} \hat{n}_{i, \uparrow} \hat{n}_{i \downarrow},
\end{equation}
where $\hat{c}^\dagger_{i, \sigma}$ ($\hat{c}_{i, \sigma}$) is the creation (annihilation) 
operator of an electron at site $i$ with spin $\sigma$, 
$\hat{n}_{i, \sigma} = \hat{c}^\dagger_{i, \sigma} \hat{c}_{i, \sigma}$ is the electron 
number operator, $t$ is the hopping integral, and $U$ is the on-site Coulomb interaction.  
$\langle i, j \rangle$ represents the nearest-neighbor pair of sites.  
We restrict ourselves to the cases of the one-dimensional chain and two-dimensional 
square lattice at half filling. 

%%%%%%%%%%%%%%%
%\section{A brief review of typical pure quantum state}
%%%%%%%%%%%%%%%

To construct the grand-canonical TPQ state, we prepare a random vector
\begin{equation}
  | \psi_{0} \rangle = \sum_{x = 1}^{N_{\mathrm{F}}} c_x | x \rangle,
\end{equation}
where $| x \rangle$ is the orthogonal basis of the Hamiltonian, 
$N_{\mathrm{F}} = \sum_{n=0}^{2L_\mathrm{s}} \frac{\left( 2L_\mathrm{s} \right)!}{n! \left( 2L_\mathrm{s}-n \right)!} $ 
is the dimension of the Fock space, $L_\mathrm{s}$ is the number of lattice sites in the system, 
and $c_x$ is a random complex number satisfying $\sum_{x} \left| c_x \right|^2 = 1$. 
This state corresponds to the TPQ state at $\beta = 0$ \cite{Imada1986JPSJ}, 
so that the average value at infinite temperature is given by 
$\langle \hat{O} \rangle^{\mathrm{ens}}_{\beta = 0} = \overline{\langle \psi_{0} | \hat{O} | \psi_{0} \rangle}$, 
where $\overline{\cdots}$ is the average over the random samplings. 

The unnormalized grand-canonical TPQ state at temperature $\beta$ and 
chemical potential $\mu$ is given as \cite{Hyuga2014PRB}
\begin{equation}
 | \psi_{\beta, \mu} \rangle
 = e^{-\beta \hat{\mathcal{K}} / 2} | \psi_0 \rangle
\end{equation}
with $\hat{\mathcal{K}} = \hat{\mathcal{H}} - \mu \sum_{i, \sigma} \hat{n}_{i, \sigma}$.  
Then, instead of Eq.~(\ref{eq:grand_canonical_ens}), we obtain the ensemble average of 
the operator $\hat{O}$ as 
\begin{equation}
 \langle \hat{O} \rangle^{\mathrm{ens}}_{\beta, \mu}
	= \frac{\overline{\langle \psi_{\beta, \mu} | \hat{O} | \psi_{\beta, \mu} \rangle}}{\overline{\langle \psi_{\beta, \mu} | \psi_{\beta, \mu} \rangle}}.  
\end{equation}
The variance from the average value
becomes exponentially small as the number of sites increases \cite{Sugiura2013PRL}, 
which means that the average value obtained from the single TPQ state is equal to 
the ensemble average in the thermodynamic limit. 
 
%%%%%%%%%%%%%%%
%\section{cluster purturbation theory}
%%%%%%%%%%%%%%%

\begin{figure}[tb]
\begin{center}
\includegraphics[width=0.95\columnwidth]{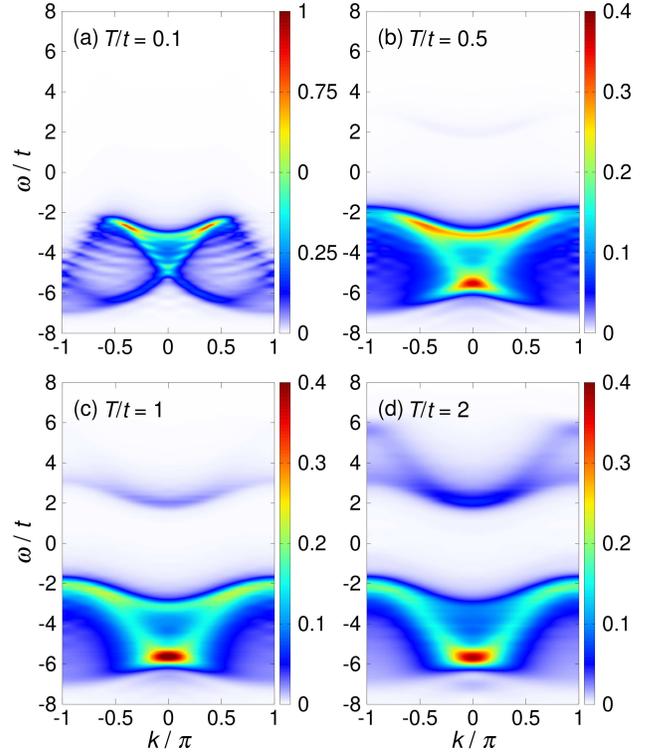}
\end{center}
\caption{Contour plot of the photoemission spectra for the one-dimensional 
Hubbard model at (a) $T/t = 0.1$, (b) $T/t=0.5$, (c) $T/t=1$, and (d) $T/t=2$ by the CPT.  
We set $U/t=8$ and assume the size of the cluster to be $L=12$ and the number 
of Lanzcos iterations to be $N_{\mathrm{L}} = 100$.  Ten random samplings for the 
TPQ-state calculations are averaged. 
}\label{fig:cpt}
\end{figure}

\begin{figure}[tb]
\begin{center}
\includegraphics[width=\columnwidth]{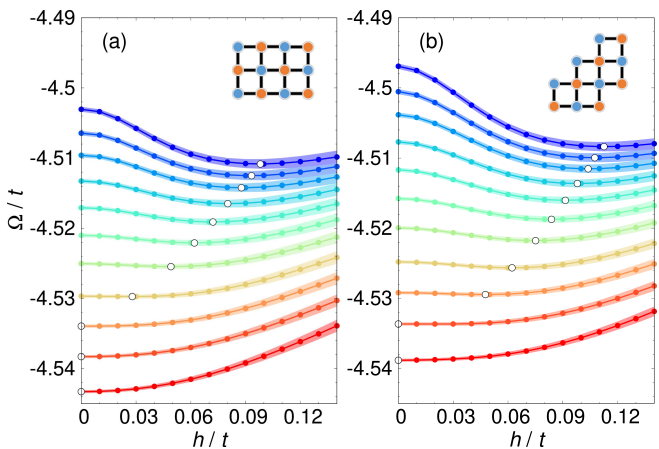}
\end{center}
\caption{Calculated grand potentials as a function of the antiferromagnetic 
Weiss field at temperatures from $T/t=0.2$ (blue line) to 0.3 (red line) in 
the interval of 0.01.  We assume the square-lattice Hubbard model at $U/t = 8$ 
and set the size of the clusters to be $L=12$ with the shape shown in the inset. 
The width of the lines indicates the variance coming from the random samplings 
of the TPQ states.  The white open circles represent the points at which the 
grand potential takes its minimum value.  The number of random samplings for 
the TPQ states is 300 for calculating $\Omega'$ and 45 for calculating 
Green's functions.  
}\label{fig:VCA}
\end{figure}

 The retarded dynamical correlation function of the operators $\hat{A}$ and
 $\hat{B}$ is given by 
 \begin{align}
 G_{\hat{A} \hat{B}} (z)
 	&= -i \int^{\infty}_0 \mathrm{d} t \, e^{i z t}
 		\langle \hat{A} (t) \hat{B} (0)
 		\mp \hat{B} (0) \hat{A} (t) \rangle^{\mathrm{ens}}_{\beta, \mu} 
 \notag \\
	&= G^{+}_{\hat{A} \hat{B}} (z) \mp G^{-}_{\hat{B} \hat{A}} (z),
\end{align}
where $\hat{A} (t) = e^{i \hat{\mathcal{K}} t} \hat{A} e^{-i \hat{\mathcal{K}} t}$ is 
the Heisenberg representation, and the upper (lower) sign is for bosonic (fermionic) operators. 
This function is calculated by the Lanczos procedure \cite{Jaklic1994PRB, Jaklic2000AP, 
Aichhorn2003PRB}.  The single-particle Green's function may then be written as 
$G_{\hat{c}_{i, a}, \hat{c}^\dagger_{j,b}}$, 
where $a$ and $b$ represent the internal degrees of freedom of the system, 
such as spin, orbital, sublattice, etc. We abbreviate the single-particle Green's function 
as $G_{ia, jb}$.

To calculate the single-particle excitation spectra in the thermodynamic limit, 
we use the CPT, where the original lattice is divided into the equal-shape disconnected 
clusters of $L$ sites.  We call the collection of the clusters forming the superlattice 
the reference system, and we denote $\tilde{\bm{k}}$ as the wave vector in the 
Brillouin zone defined for the superlattice, i.e., the wave vector is written as 
$\bm{k} = \tilde{\bm{G}} + \tilde{\bm{k}}$, where $\tilde{\bm{G}}$ is the reciprocal 
lattice vector for the superlattice.  The Green's function in the CPT is given 
by \cite{Senechal2000PRL, Senechal2002PRB}
\begin{equation}
 G^{\mathrm{CPT}}_{a, b} (\bm{k}, z)
	= \frac{1}{L} \sum_{i, j}^{L} e^{- i \bm{k} \cdot \left( \bm{r}_i - \bm{r}_j \right) } \tilde{G}_{ia, jb} (\tilde{\bm{k}}, z)
\end{equation}
where $\bm{r}_i$ and $\bm{r}_j$ are the site positions in the cluster and
\begin{equation}
  \tilde{G} (\tilde{\bm{k}}, z) = \left[ \left( {G'} (z) \right)^{-1} - V (\tilde{\bm{k}}) \right]^{-1},
\end{equation}
is the approximate single-particle Green's function of the original system in the matrix 
representation.  Here, ${G'}^{-1} (z)$ is the inverse of the exact Green's function of 
the reference system and $V (\tilde{\bm{k}})$ is the Fourier transform of 
$V = \left( {G'}^{(0)} \right)^{-1} - \left( G^{(0)} \right)^{-1}$, 
the difference between the one-body-term Green's functions 
of the reference and original systems.  
With the CPT Green's function, the single-particle spectral function is given by 
$A (\bm{k}, \omega) = -\frac{1}{\pi} \mathrm{Im} \, \mathrm{Tr} \, G^{\mathrm{CPT}} (\bm{k}, z = \omega + i \eta)$.

First, to demonstrate the validity of the TPQ-state approach combined with the CPT, 
we calculate the photoemission spectral function $A^{-}(\bm{k}, \omega)$ for 
the one-dimensional Hubbard model at half filling at various temperatures.  
The results are shown in Fig.~\ref{fig:cpt}, which may be compared with the results 
obtained by the finite-temperature time-dependent density-matrix renormalization 
group method \cite{Nocera2019PRB}.  We find that the agreement is very good: 
At low temperatures, the spinon and holon dispersions caused by the spin-charge 
separation \cite{Kim1997PRB} appear, and with increasing temperature, 
the spinon dispersion gradually disappears while the holon dispersion remains. 
Also noticed is that thermally excited electrons go into the upper Hubbard band, 
leading to the emergence of the spectral weight above the Fermi level. 
We thus conclude that the TPQ-state approach combined with the CPT indeed 
works very well.  

%%%%%%%%%%%%%%%
%\subsection{Variational cluster approach}
%%%%%%%%%%%%%%%

\begin{figure}[tb]
\begin{center}
\includegraphics[width=\columnwidth]{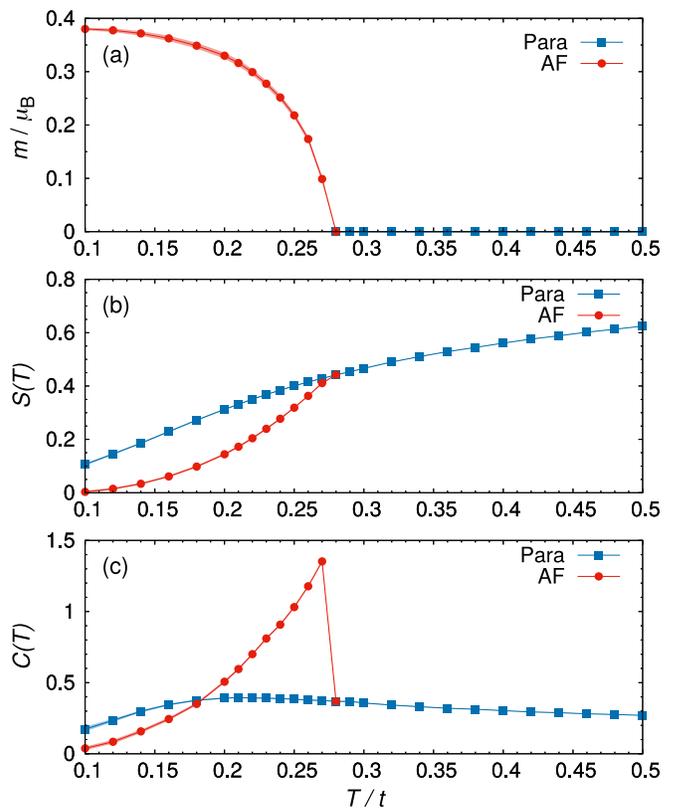}
\end{center}
\caption{(a) Calculated order parameter $m$, entropy $S$, and specific heat $C$ 
as a function of temperature $T$ for the paramagnetic (Para) and antiferromagnetic 
(AF) phases of the square-lattice Hubbard model at $U/t = 8$.  We assume the 
cluster of the size $L = 12$ with the shape shown in the inset of Fig.~\ref{fig:VCA}(a).  
}\label{fig:C}
\end{figure}

\begin{figure*}[tb]
\begin{center}
\includegraphics[width=\linewidth]{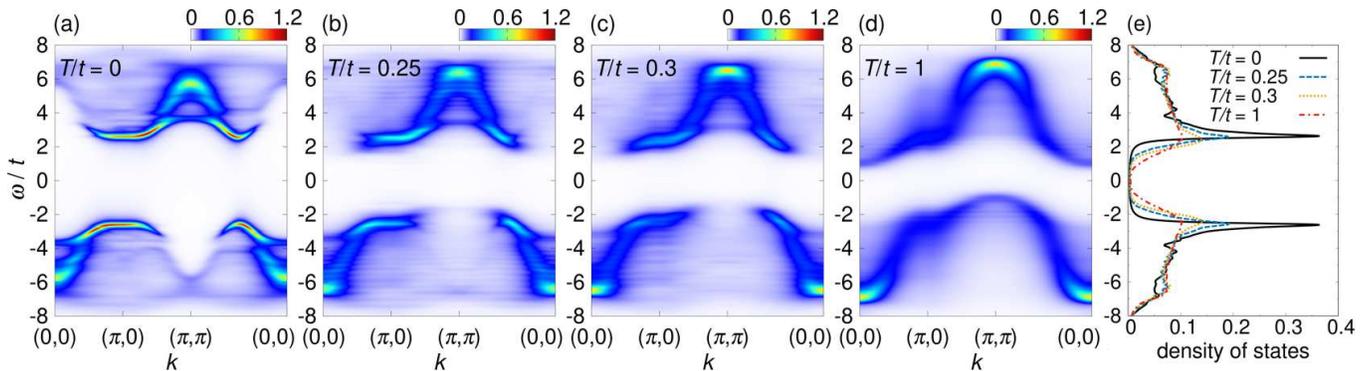}
\end{center}
\caption{Contour plot of the calculated single-particle excitation spectra $A(\bm{k},\omega)$ 
for the square-lattice Hubbard model at $U/t = 8$ both below and above the antiferromagnetic 
transition temperature $T_{\mathrm{N}}/t = 0.28$. 
The results are given at (a) $T/t=0$, (b) $T/t=0.25$, (c) $T/t=0.3$, and (d) $T/t=1$.  
Calculated densities of states at different temperatures are also shown in (e).  
The cluster size is set to be $L=12$ with the shape shown in the inset of 
Fig.~\ref{fig:VCA}(a).  Ten random samplings for the TPQ-state calculations are averaged. 
}\label{fig:SPES}
\end{figure*}

Next, we discuss the TPQ-state approach combined with the VCA.  
The VCA is an extended version of the CPT, where the grand potential 
of the original system is minimized by varying the self-energy of the 
reference system, or equivalently by varying one-body parameters in 
the cluster.  The grand potential (per site) is given by \cite{Potthoff2003PRL}
\begin{equation}
 \Omega (\beta) = \Omega' (\beta) - \frac{1}{\beta L N_{\mathrm{c}}} 
\sum_{\tilde{\bm{k}}} \sum_n \ln \det \left[ 1 - V (\tilde{\bm{k}}) G' (i \varepsilon_n) \right],
\label{eq:grand_potential}
\end{equation}
where $N_{\mathrm{c}}$ is the number of the clusters in the system, 
$\varepsilon_n = (2n+1) \pi / \beta$ is the Matsubara frequency, and
\begin{equation}
 \Omega' (\beta) = - \frac{1}{\beta L} \ln \left( N_{\mathrm{F}} 
\overline{\langle \psi_{\beta, \mu} | \psi_{\beta, \mu} \rangle} \right)
\end{equation}
is the grand potential (per site) of the reference system.  
Here, $| \psi_{\beta, \mu} \rangle$ is the TPQ state in the cluster. 
Also, the average value of an operator $\hat{O}$ in the VCA is given by 
\begin{equation}
 \langle \hat{O} \rangle
	= \frac{1}{L N_{\mathrm{c}}} \sum_{\tilde{\bm{k}}} \sum_n \mathrm{Tr} \left\{ \hat{O} G' (i \varepsilon_n) \left[ 1 - V (\tilde{\bm{k}}) G' (i \varepsilon_n) \right] \right\}.
\end{equation}
Because Eq.~(\ref{eq:grand_potential}) contains the summation over an infinite 
number of Matsubara frequencies, which is difficult to calculate directly, we 
replace the summation in the high frequency part by the contour integral 
in the complex plane, as in Ref.~\onlinecite{Seki2018PRB}.  

The spontaneous symmetry breaking for antiferromagnetism is treated within 
the framework of the VCA by adding the Weiss field 
$\hat{\mathcal{H}}' = h \sum_{i, \sigma} \sigma e^{i \bm{Q} \cdot \bm{r}_i} \hat{n}_{i, \sigma}$ 
to the Hamiltonian of the reference system, where $h$ is the strength of the Weiss field 
and $\bm{Q}$ is the ordering vector.  The magnetic moment 
$m = \frac{1}{L} \mu_{\mathrm{B}} \sum_{i, \sigma} \sigma e^{i \bm{Q} \cdot \bm{r}_i} \langle \hat{n}_{i, \sigma} \rangle$, 
where $\mu_{\mathrm{B}}$ is the Bohr magneton, is the order parameter of this system.
Figures \ref{fig:VCA}(a) and \ref{fig:VCA}(b) show the grand potentials as a function of 
the Weiss field for antiferromagnetism with $\bm{Q} = (\pi, \pi)$, calculated for the 
square-lattice Hubbard model at half filling with $U/t=8$.  We assume two different 
clusters of size $L=12$ shown in the inset of Figs.~\ref{fig:VCA}(a) and \ref{fig:VCA}(b).  
We find that the grand potentials have a minimum at a finite Weiss field, 
indicating that the antiferromagnetic ordering actually occurs.  The calculated 
magnetic moment as a function of temperature is shown in Fig.~\ref{fig:C}(a), where 
we find that the phase transition is of the second order with the transition 
temperature $T_{\mathrm{N}}/t = 0.28$.  
We should note here that, according to the Mermin-Wagner-Hohenberg theorem 
\cite{Mermin1966PRL, Hohenberg1967PR}, any spontaneous breaking of continuous symmetries 
is prohibited in two dimension at finite temperature.  Our results contradict with this theorem.  
This is because the clustering of the original lattice cuts off the long-range correlations 
and correlations beyond the cluster size are treated only in a mean-field level where the Weiss 
field explicitly breaks the symmetry.  Reflecting this situation, the transition temperature 
decreases as the cluster size increases \cite{Seki2018PRB}. 

An advantage of the VCA is a direct calculation of the thermodynamic function 
$\Omega(T)$, which enables one to calculate the entropy and specific heat of the 
system directly as 
$S (T) = - (\mathrm{d}\Omega/\mathrm{d}T)$ and $C(T)=T(\mathrm{d}S/\mathrm{d}T)$.  
Figures \ref{fig:C}(b) and \ref{fig:C}(c) show the results for the temperature dependence 
of the entropy and specific heat of the system, respectively, calculated for the 
square-lattice Hubbard model at half filling with $U/t=8$.  We observe that a rapid 
decrease in the entropy occurs below the transition temperature, which is associated 
with a jump in the specific heat.  These behaviors are consistent with the results given 
in Ref.~\onlinecite{Seki2018PRB}, ensuring that the TPQ-state approach combined 
with the VCA is a powerful tool for calculating the thermodynamic quantities in systems 
where any phase transition occurs.  

Finally, we calculate the temperature dependence of the single-particle excitation 
spectra in the TPQ-state approach combined with the VCA, assuming the same 
model as above.  The results are shown in Figs.~\ref{fig:SPES}(a)-(d), 
together with the density of states in Fig.~\ref{fig:SPES}(e).  
We find that the energy gap opens both below and above the transition temperature, 
so that the system is insulating in the entire temperature range.  The band dispersions 
do not change largely across the phase transition point 
[see Figs.~\ref{fig:SPES}(b) and \ref{fig:SPES}(c)].  
We also find that the energy gap gradually decreases at very high temperatures 
because many doublon states are created by thermal excitations.  
These results demonstrate the behavior of the Mott gap over a wide temperature range.  

%%%%%%%%%%%%%%%%%%%
%\subsection{Summary}
%%%%%%%%%%%%%%%%%%%

In summary, based on the typicality, we developed the TPQ-state approach combined 
with the CPT and VCA and calculated the single-particle excitation spectra and specific 
heat of the Hubbard model at finite temperatures successfully.  
The primary advantage of the present approach is that the expectation value of the physical 
quantity can be obtained from the single TPQ state, so that one can efficiently investigate 
the thermodynamic properties of the system at high temperatures without diagonalizing 
the large Hamiltonian matrix.  We note that the present approach is not restricted to the 
Hubbard model but is also applicable to other fermionic as well as bosonic models with 
strong correlations, and may also be useful for frustrated quantum systems where many 
nearly-degenerate eigenstates exist at low energies.  Moreover, the calculations of the 
two-body correlation functions by the CPT and VCA \cite{Brehm2010EPL, Raum1908arXiv} 
can be extended to finite temperatures using the present TPQ-state approach.  
Our studies thus pave the way for future investigations of strongly correlated quantum 
systems at finite temperatures.  

%\section*{Acknowledgments}

We thank A.~Sugita for tutorial lectures on the typicality-based approaches and 
T.~Yamaguchi for enlightening discussions.  This work was supported in part by 
JSPS KAKENHI Grants (Nos.~JP19J20768, JP19K14644, and JP17K05530) and 
Keio University Academic Development Funds for Individual Research. 
The numerical calculations were carried out on XC40 at YITP in Kyoto University.


\begin{thebibliography}{99}

%Quantum cluster approximation
\bibitem{Maier2005RMP} T. Maier, M. Jarrell, T. Pruschke, and M. H. Hettler, Rev. Mod. Phys. \textbf{77}, 1027 (2005).
\bibitem{Senechal0806arXiv} D. S\'{e}n\'{e}chal, arXiv:0806.2690

% Reviews of DMFT with QMC
\bibitem{Georges1996RMP} A. Georges, G. Kotliar, W. Krauth, and M. J. Rozenberg, Rev. Mod. Phys. \textbf{68}, 13 (1996).
\bibitem{Gull2011RMP} E. Gull, A. J. Millis, A. I. Lichtenstein, A. N. Rubtsov, M. Troyer, and P. Werner, Rev. Mod. Phys. \textbf{83}, 349 (2011).

% negative sign problem in QMC method
\bibitem{Loh1990PRB} E. Y. Loh, Jr., J. E. Gubernatis, R. T. Scalettar, S. R. White, D. J. Scalapino, and R. L. Sugar, Phys. Rev. B \textbf{41}, 9301 (1990).
\bibitem{Troyer2005PRL} M. Troyer and U.-J. Wiese, Phys. Rev. Lett. \textbf{94}, 170201 (2005).

%TPQ state
\bibitem{Sugiura2012PRL} S. Sugiura and A. Shimizu, Phys. Rev. Lett. \textbf{108}, 240401 (2012).
\bibitem{Sugiura2013PRL} S. Sugiura and A. Shimizu, Phys. Rev. Lett. \textbf{111}, 010401 (2013).
\bibitem{Hyuga2014PRB} M. Hyuga, S. Sugiura, K. Sakai, and A. Shimizu, Phys. Rev. B \textbf{90}, 121110(R) (2014).

\bibitem{Goldstein2006PRL} S. Goldstein, J. L. Lebowitz, R. Tumulka, and N. Zangh\`{i}, Phys. Rev. Lett. \textbf{96}, 050403 (2006).
\bibitem{Popescu2006NP} S. Popescu, A. J. Short, and A. Winter, Nat. Phys. \textbf{2}, 754 (2006).
\bibitem{Sugita2007NPCS} A. Sugita, Nonlinear Phenom. Complex Syst.\textbf{10}, 192 (2007).
\bibitem{Reimann2007PRL} P. Reimann, Phys. Rev. Lett. \textbf{99}, 160404 (2007).

\bibitem{Imada1986JPSJ} M. Imada, and M. Takahashi, J. Phys. Soc. Jpn. \textbf{55}, 3354 (1986).
\bibitem{Hams2000PRE} A. Hams and H. De Raedt, Phys. Rev. E \textbf{62}, 4365 (2000). 

% finite temperature Lanczos method
\bibitem{Jaklic1994PRB} J. Jakli\v{c} and P. Prelov\v{s}ek, Phys. Rev. B \textbf{49}, 5065 (1994).
\bibitem{Jaklic2000AP} J. Jakli\v{c} and P. Prelov\v{s}ek, Adv. Phys. \textbf{49}, 1 (2000).
\bibitem{Aichhorn2003PRB} M. Aichhorn, M. Daghofer, H. G. Evertz, and W. von der Linden, Phys. Rev. B \textbf{67}, 161103(R) (2003). 

%Dynamical TPQ
\bibitem{Bartsch2009PRL} C. Bartsch and J. Gemmer, Phys. Rev. Lett. \textbf{102}, 110403 (2009).
\bibitem{Elsayed2013PRL} T. A. Elsayed and B. V. Fine, Phys. Rev. Lett. \textbf{110}, 070404 (2013).
\bibitem{Steinigeweg2014PRL} R. Steinigeweg, J. Gemmer, and W. Brenig, Phys. Rev. Lett. \textbf{112}, 120601 (2014).
\bibitem{Monnai2014JPSJ} T. Monnai and A. Sugita, J. Phys. Soc. Jpn. \textbf{83}, 094001 (2014).
\bibitem{Jin2015PRB} F. Jin, R. Steinigeweg, F. Heidrich-Meisner, K. Michielsen, and H. De Raedt, Phys. Rev. B \textbf{92}, 205103 (2015).
\bibitem{Steinigeweg2016PRB} R. Steinigeweg, J. Herbrych, F. Pollmann, and W. Brenig, Phys. Rev. B \textbf{94}, 180401(R) (2016).
\bibitem{Yamaji2018arXiv} Y. Yamaji, T. Suzuki, M. Kawamura, arXiv:1802.02854.
\bibitem{Richter2019PRB} J. Richter and R. Steinigeweg, Phys. Rev. B \textbf{99}, 094419 (2019).

%CPT
\bibitem{Senechal2000PRL} D. S\'{e}n\'{e}chal, D. Perez, and M. Pioro-Ladri\`{e}re, Phys. Rev. Lett. \textbf{84}, 522 (2000).
\bibitem{Senechal2002PRB} D. S\'{e}n\'{e}chal, D. Perez, and D. Plouffe, Phys. Rev. B \textbf{66}, 075129 (2002).

%VCA
\bibitem{Potthoff2003EPJB} M. Potthoff, Eur. Phys. J. B \textbf{32}, 429 (2003).
\bibitem{Potthoff2003EPJB2} M. Potthoff, Eur. Phys. J. B \textbf{36}, 335 (2003).
\bibitem{Potthoff2003PRL} M. Potthoff, M. Aichhorn, and C. Dahnken, Phys. Rev. Lett. \textbf{91}, 206402 (2003).
\bibitem{Dahnken2004PRB} C. Dahnken, M. Aichhorn, W. Hanke, E. Arrigoni, and M. Potthoff, Phys. Rev. B \textbf{70}, 245110 (2004).
\bibitem{Seki2018PRB} K. Seki, T. Shirakawa, and S. Yunoki, Phys. Rev. B \textbf{98}, 205114 (2018).

%tDMRG
\bibitem{Nocera2019PRB} A. Nocera, F. H. L. Essler, and A. E. Feiguin, Phys. Rev. B \textbf{97}, 045146 (2018).
\bibitem{Kim1997PRB} C. Kim, Z.-X. Shen, N. Motoyama, H. Eisaki, S. Uchida, T. Tohyama, and S. Maekawa, Phys. Rev. B \textbf{56}, 15589 (1997).

%Marmin-Wagner-Hoenberg theorem
\bibitem{Mermin1966PRL} N. D. Mermin and H. Wagner, Phys. Rev. Lett. \textbf{17}, 1133 (1966).
\bibitem{Hohenberg1967PR} P. C. Hohenberg, Phys. Rev. \textbf{158}, 383 (1967).

%two-body correlation function by CPT and VCA
\bibitem{Brehm2010EPL} S. Brehm, E. Arrigoni, M. Aichhorn,  and W. Hanke, EPL \textbf{89}, 27005 (2010).
\bibitem{Raum1908arXiv} P. T. Raum, G. Alvarez, T. Maier, and V. W. Scarola, arXiv:1908.10361.

\end{thebibliography}
\end{document}